\documentstyle[twoside,fleqn,espcrc2,epsfig]{article}

\newcommand{\AmS}{{\protect\the\textfont2
  A\kern-.1667em\lower.5ex\hbox{M}\kern-.125emS}}

\title{The Periods Discovered by RXTE in Thermonuclear Flash Bursts}

\author{T. E. Strohmayer\thanks{also University Space Research
Association}, J. H. Swank, and W. Zhang \address{Code 662, Goddard Space
Flight Center\\Greenbelt, MD 20771, U. S. A.}}

\begin{document}

\begin{abstract}
Oscillations in the X-ray flux of thermonuclear X-ray bursts have been observed
with RXTE from at least 6 low-mass binaries, at frequencies from 330 Hz
to 589 Hz.  There appear to be preferred relations between the frequencies 
present during the bursts and those seen in the persistent flux. The amplitude
of the oscillations can exceed 50 \% near burst onset.  Except for a 
systematic increase in oscillation frequency as the burst progresses,
the frequency is stable.
Time resolved spectra track increases in the X-ray emitting area due to
propagation of the burning front over the neutron star surface, as well as
radiation driven expansion of the photosphere. The neutron star mass, radius,
and distance can be inferred when spectra are compared to theoretical
expectations. 

\end{abstract}

\maketitle

\section{OSCILLATIONS IN BURST FLUX}

Type I (thermonuclear) X-ray bursts from six bursters have been found to have
episodic, strong oscillations, without any significant evidence of harmonics.
For sources
from which multiple bursts have been observed, not all bursts have
exhibited the oscillations. The oscillations were first seen in a burst
from 4U 1728-34 \cite{Stroh96}. In an observation of 200 ksec
duration when the source was bursting, 12 bursts were seen, 6 of them
exhibiting oscillations at about 363 Hz. So far, when a source has
multiple
bursts with oscillations, the frequencies are almost the same, so that a
frequency identifies the source. It is likely that some bursts from many
other bursters also exhibit oscillations. In most cases, we have no way,
yet, of knowing whether a burster is burst active, so that we have not
been able to choose to observe it at such a time. We also do not know for
sure why sometimes the oscillations are not seen, although this seems to be
correlated with the strength of the bursts and whether they result in
radius expansion of the neutron star.  

\begin{figure}[htb]
\parbox{73mm}{\epsfig{angle=90.,file=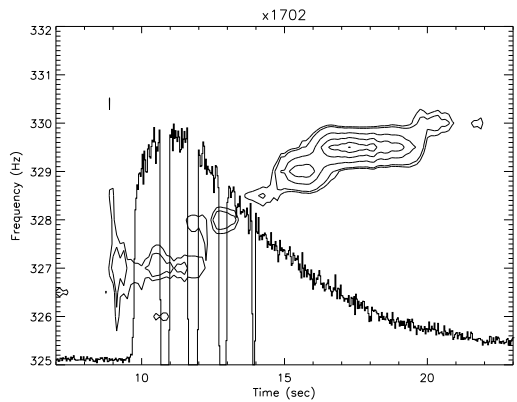,
width=73mm,height=54mm}}
\caption{Oscillations during a burst from 4U 1702-43.
Contours of Fourier power 
are plotted as a function of frequency and time. The burst
light curve is overlaid. The gaps result from insufficient telemetry
bandwidth at high flux.} 
\label{fig:largenenough}
\end{figure}

To display properties of the burst oscillations it is convenient to compute the
power spectra of 2 s intervals with a new interval sampled every 1/8 s. 
Figure 1 shows an example of a burst with a typical
appearance of the oscillations during the course of a burst, with a
slight rise in frequency of about 2 Hz to a value that stays approximately
constant during the decline of the burst. Sometimes only a portion of this kind
of pattern can be seen. Even with the change of frequency in some bursts,
$\frac{\Delta\nu}{\nu} \ge 300$. In the tails of the bursts, it can be
$\ge 1000$.

\begin{figure*}[htb]
\parbox{157mm}{\epsfig{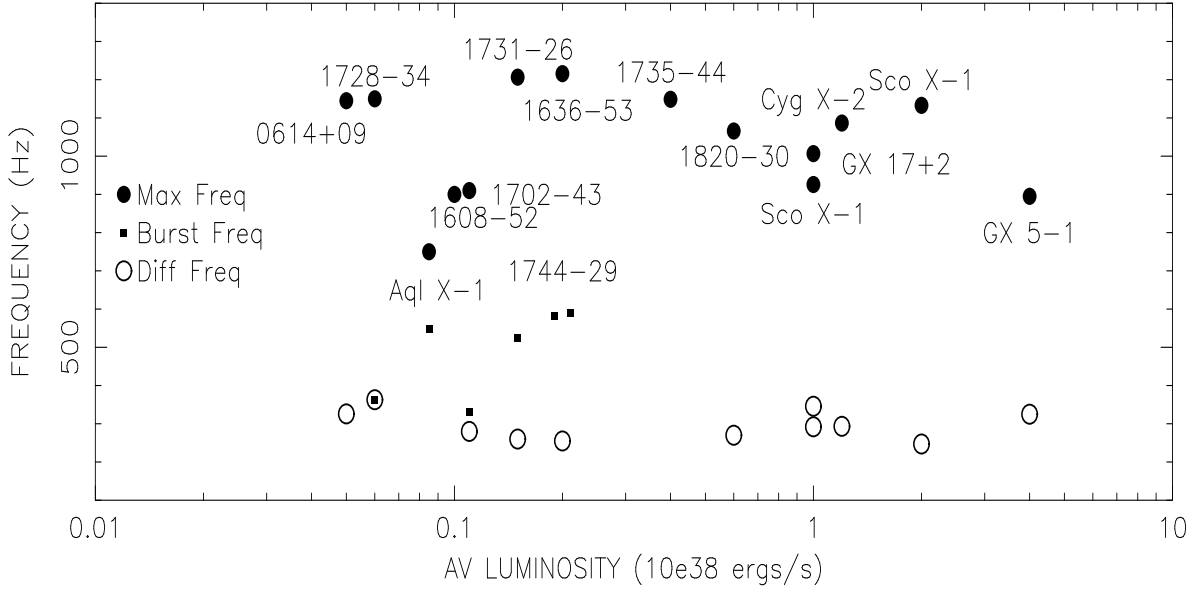}}
\caption{Kilohertz oscillations in LMXB detected with RXTE. 
Filled circles are reported
maximum QPO frequencies in the persistent emission. Open circles are 
the
difference between the two persistent emission frequencies when both are
present. Filled squares are the frequencies seen in bursts.
See van der Klis, this volume, for the most recent review of persistent emission 
results and references.}
\label{fig:largenenough} \end{figure*}

Among the six bursters with burst oscillations four also exhibit twin 
``kilohertz'' quasiperiodic oscillations (QPO), which drift in frequency with
inferred mass accretion rate during an observation.
Table~\ref{tab:periods} gives their burst oscillation periods and Figure 2
summarizes the
relations between oscillation frequencies that have been reported to
date. The frequencies are plotted with respect to estimated persistent
luminosity \cite{Christian97}, to emphasize that they are approximately
independent of it. 
For 4U 1728-34 and 4U 1702-43 the difference frequencies of the twin kilohertz
peaks are  close to the
burst frequencies. For 4U 1636-53 and KS 1731-44, the difference
frequencies
are closer to half the burst frequencies. For Aql X-1 and a burster in the
galactic center, X1743-29, possibly the same as a burster seen
previously (MXB 1743-29) \cite{StrohJahoda}, it has not been possible to
date to identify two oscillation frequencies in the persistent emission from
which to compute a difference. 

\begin{table}[hbt]
\caption{Burst Oscillation Periods}
\label{tab:periods}
\begin{tabular}{lrr}
\hline
Source          &Period(ms) &$\#$ of Bursts   	\\
\hline
4U1636-53	& $1.72$    & $4$		\\
4U 1702-43	& $3.03$    & $1$		\\
4U1728-34	& $2.76$    & $16$		\\
KS 1731-26	& $1.90$    & $1$		\\
X 1743-29	& $1.70$    & $3$		\\
Aql X-1 	& $1.82$    & $1$		\\
\hline
\end{tabular}
\end{table}

Neither of the variable persistent emission QPO frequencies is a candidate for
the rotation frequency of the neutron star, but the difference
frequencies have been linked to the spin frequency \cite{Stroh96,White97}
despite the
variations in difference frequencies seen for Sco X-1 and 4U 1608-52. A strong
case has been made that the burst frequencies are directly related to the 
stellar spin frequency \cite{Stroh98}. 
For the burst oscillations the variations in frequency for a given source are
rather small and explanations appear plausible, as we will discuss. In
either case a model in terms of rotation of a neutron star with a
non-uniform temperature distribution appears reasonable. There could be a
single hot (or cold) spot giving rise to rotational modulation or perhaps a
spot associated
with each magnetic pole, even if the magnetic fields are much smaller
than the $10^{12}$ G fields of high mass binary accreting neutron stars.
In this paper we discuss the
evidence that the burst frequencies are the rotation frequencies of the
neutron stars in low mass X-ray binaries (LMXB). 

Other explanations might have been, a priori possible. When
a neutron star is hot
because of a thermonuclear flash, oscillations in the inner part of
a remnant disk, not blown back by the explosion, could obscure part of the
glowing star periodically. We will see, however, that there are associated
oscillations in the temperature, rather than the area, making this
explanation less probable. The explosion and the transfer of heat to the
surface, by a combination of diffusion and convection, could excite
oscillation modes of the neutron stars. Such modes have been studied
theoretically
\cite{Stroh92,Bildsten98} and some have frequencies in the observed
range.
However, there is no compelling reason why the frequency spectrum of the 
modes would be so
simple and estimates of the amount of energy that would be in these
oscillations are a small fraction of the amplitudes actually observed.

We report on our studies of properties of the burst
oscillations that tend to confirm interpretation in terms of rotation:
the sinusoidal pulse shape, the very high amplitude which can occur at the
onset of bursts, the correlation with a temperature oscillation, and
coherence of the oscillations through the bursts. We have
expected theoretically that these neutron stars are rotating fast and the 
large population of millisecond radio pulsars has implied the existence of
progenitors such as LMXB,
so that the theoretical implication that we should find such
objects is very strong. 
Finally, we address briefly the neutron star properties that the burst
observations can strongly constrain.

\section{BURST OSCILLATION PROPERTIES}

\subsection{Increasing Area during the Rise}

The data allow the pulsed amplitude to be tracked during bursts as a
function of energy \cite{Stroh97}.
The spectra of the bursts during the rise can be fit every 0.125 s and    
yield a color temperature and an apparent radius to better than 10 \%
during a period when the apparent radius changes by $>$ 50 \%. 
Expected corrections to the spectra would make the scale of radius increase
larger.

In some bursts the oscillations are
seen very close to the beginning of the rise in flux, within the
first 100 ms. Within such time intervals, the oscillations are coherent
and can be folded on the period. Fractional amplitudes near and exceeding
50 \% are
seen both in bursts 4U 1728-34, in which the difference frequency of
363 Hz is about the burst frequency, and in bursts from 4U 1636-53, in
which
the difference frequency of about 255 Hz is closer to half the burst
frequency of 581 Hz. The fractional amplitude decreases strongly as the
burst flux rises to the peak. This suggests that the thermonuclear
burning starts as a localized hot spot on the neutron star and from there
spreads around the neutron star. Calculations have suggested the burning is
inhomogeneous \cite{Bildsten95}. Simulations of emission from the surface
of a neutron star show that this model can
reproduce the observed behavior of the fractional amplitude.

Because the neutron star gravitational field bends light around it the 
intensity of radiation emitted from a surface hot spot on the neutron star
depends on the rotational phase of the spot and the compactness, $M/R$ of the
neutron star. Here $M$ and $R$ are the stellar mass and radius, respectively.
This is taken into account in simulations of the expected pulse growth and 
decay \cite{Stroh97}. The more compact the neutron
star, that is, the higher is M/R, the more severe is the bending of light,
the more smeared the pulse, and the lower would be the modulation of
radiation from the surface
\cite {Stroh98}. Thus the high level of modulation of the flux that is
observed puts a strong constraint on the compactness of the neutron star.
The neutron stars must be fairly large. If the intensity is actually from
opposing poles of a neutron star, these constraints become almost
impossible for known and tested equations of state of nuclear matter. This
suggests that the hypothesis of two antipodal spots may be untenable.

We have found no substantial differences between the 
behavior of the burst oscillations for 4U 1728-34 and 4U 1636-53 to 
contradict the hypothesis that the mechanisms are the same.
The similarity of the burst rise behavior thus argues in favor of both  
being the neutron star spin.

Interpretation of the oscillations at the beginning of a burst in terms of
a spreading hot spot raises the question of what mechanism causes the 
oscillations that
appear during the decays of bursts. There would have to be a remnant or
recurring inhomogeneity on the surface of the neutron star. Possibly what
was hot first is now 
a cooler spot, although conductivity might tend to make the star
homogeneous. So far at
least, there is no evidence there are different numbers of hot spots
arising on the neutron star. This suggests that the occurrence is not
random and a magnetic pole is an obvious candidate. 

\begin{figure}[htbp]
\parbox{73mm}{\epsfig{angle=0.,file=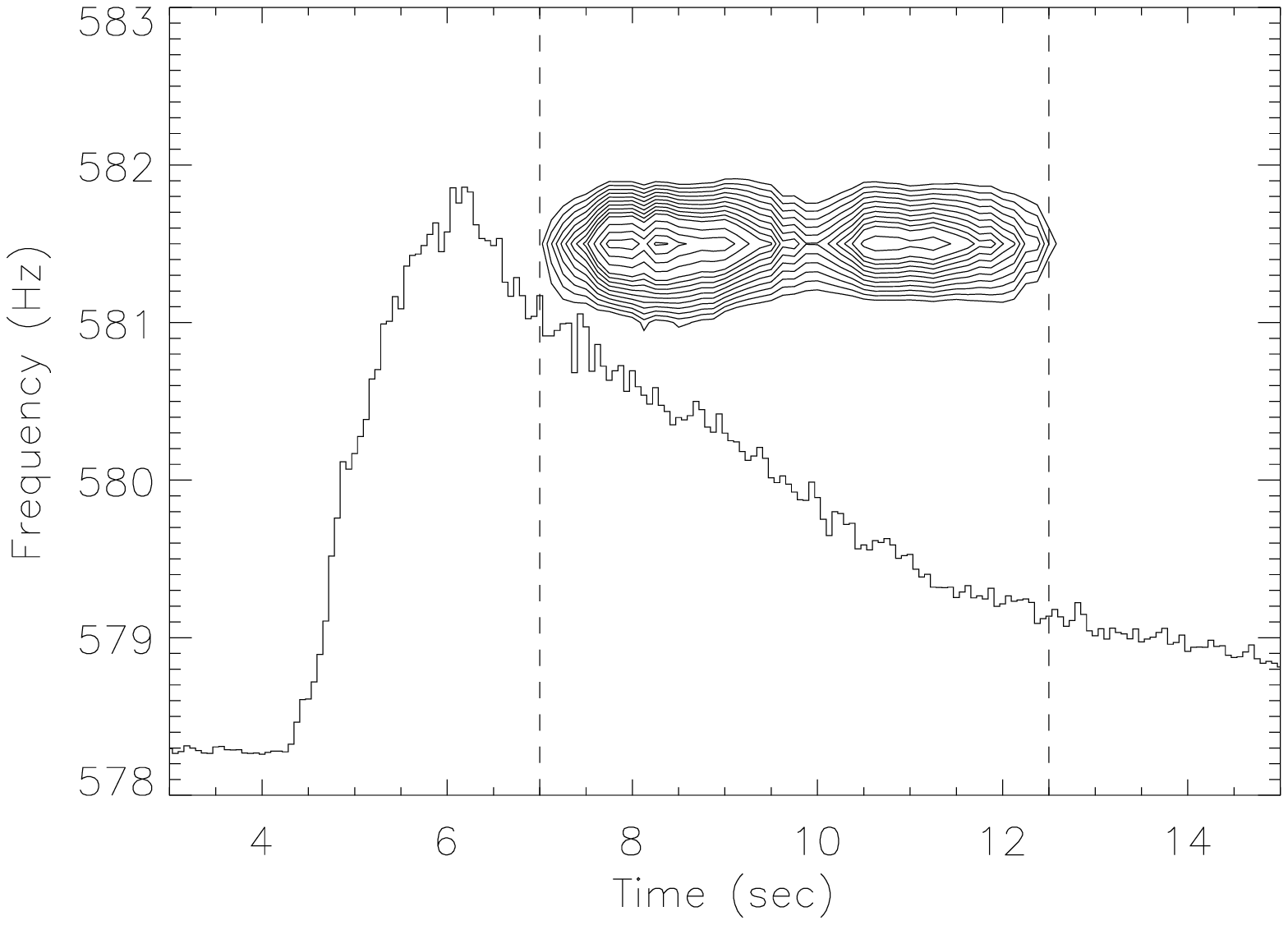
,width=73mm,height=54mm}}

\parbox{73mm}{\epsfig{angle=0.,file=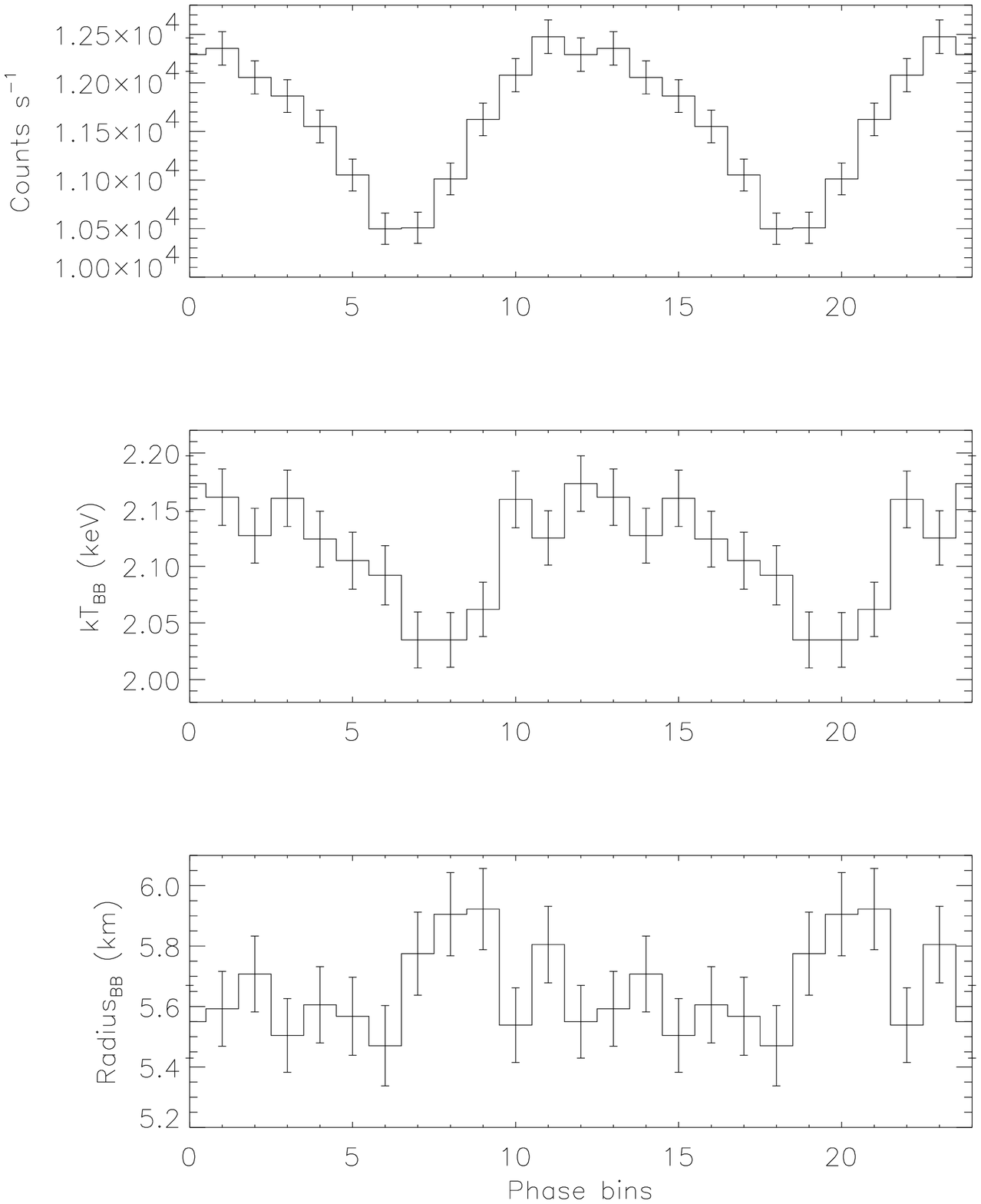
,height=84mm,width=73mm}}
\caption{Oscillations in a burst observed from 4U 1636-54 by I. Lapidus.
Contours of Fourier 
power with the burst profile overlaid (top). Epoch folded pulse profile,
blackbody kT, and apparent
radius (bottom). Note the 6 \% change in the blackbody temperature with 
pulse phase associated with the 20 \% amplitude in luminosity. A small asymmetry
due to Doppler shifts would be expected and a
confirmed measurement would be important.}
\end{figure}

\subsection{Temperature versus Area Oscillations}

In cases where the duration of coherent oscillations is long enough and
the amplitude strong enough, 
pulse phase spectroscopy is possible. In a burst from 4U 1636-53 shown in
Figure 3, the pulsations for 5 s in the burst decay produce a 6 \% modulation in
the derived 
blackbody kT.

\subsection{Coherence of Burst Oscillations}

For the burst oscillations to represent the rotation period of the neutron
stars, there must be an explanation for the 1-2 Hz change in frequency observed
during bursts, since this cannot represent a real change in the angular momentum
of the entire neutron star on such a short time scale.  A model which has been
discussed \cite{StrohJahoda} is that the heated burning layer
expands on the order of 10-30 m and is not rigidly attached to the neutron
star, so that conservation of angular momentum causes the expanded hot
material to rotate more slowly than material deeper in. 
It might then wrap around the neutron star about once during the initial 
second of the burst rise. A crude estimate of the height above the surface that
the material would have to rise in order to produce such an effect is given by
equating the angular momentum of a thin shell at the stellar surface, $R$, and
at a slightly higher altitude, $R+\Delta R$, including the correction
introduced by the gravitational time dilation from different radii.

\begin{equation}
\frac {(R + \Delta R)^2}{R^2}=
\sqrt{\frac{1 - \frac{R_s}{R + \Delta R}}{1 - \frac{R_s}{R}}} 
\frac{\nu _{1,\infty}}{\nu _{0,\infty}}
\end{equation}

Here, $\nu_{0,\infty}$ and $\nu_{1,\infty}$ are the frequencies measured 
near burst onset and burst decay, respectively.
An observed frequency lower by 1 Hz than a final frequency of 500 Hz would
correspond to a 10-20 m expansion during the rising part of the burst in
comparison to the tail of the burst. 
Whatever the exact explanation, this serves to illustrate that modest changes in
the source of a modulation near the neutron star surface could cause the
amount of frequency change observed. 

In this model, gradual changes in the pulse
peak phases would be seen. Given a frequency during the tail of the burst
and a frequency at the beginning, a frequency derivative can always bridge a
gap. But for a given frequency change, the pulses need not remain in
phase. In fact, Zhang et al. \cite{Zhang98} apply the frequency derivative
seen in a burst from Aql X-1, correct the times to that of the neutron
star and find that almost all of the power in the oscillation is recovered,
consistent with coherence of the underlying clock.

\begin{figure}[htbp]
\parbox{75mm}{\epsfig{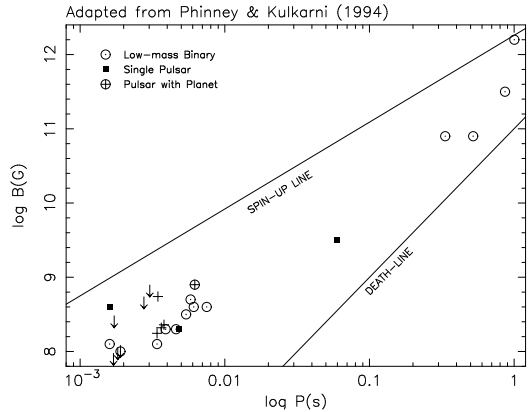}}
\caption{Recycled radio pulsar candidates and candidate burst pulsar
properties. The upper limits are derived according to White \&
Zhang (1997) for the observed burst oscillation periods. Also shown (+) 
are the limits if difference frequencies are used instead.
Evolution as radio pulsars would take them to the right.}
\label{fig:largenenough}
\end{figure}

\section{MILLISECOND PULSAR PROGENITORS}

The LMXB are part of the old population of stars in the
Galaxy. They have been accreting for a long time compared to high mass
X-ray binaries \cite{Battacharya95}. It was speculated when the low
magnetic field millisecond radio pulsars were found that they were
recycled pulsars spun up by accretion in LMXB. 
The periods we find are certainly in the approximate region of the periods of the
millisecond radio pulsars as shown in Figure 4. We do not have firm
measurements of the
magnetic fields but they have been expected to have surface fields of
$\approx 10^8 - 10^9 $ G and the current data can be interpreted
in terms of upper limits on the field of that order \cite{White97}.
However, careful estimates of the number of low field millisecond radio
pulsars \cite{Phinney94} ($5 \times 10^4$)  imply a birthrate about 10
times the birthrate of LMXB. The long time needed for
LMXB to spin up the neutron star ($8 \times 10^7 $) yr when accreting
at the observed
luminosities of only $0.01-0.1$ of the Eddington limit imply a  birth
rate insufficient to be responsible for the observed millisecond radio
pulsars. A large part of
that population  may have a different origin,
perhaps accreting low mass binaries which have a phase of
accretion so strong that they are not seen as X-ray sources
(\cite{Burderi96}), perhaps actually formed with low magnetic fields.

\section{PROPERTIES OF THE BURST NEUTRON STARS}

A radius expansion burst from 4U
1820-30 illustrates the possible information that can be obtained from the
bursts.  Spectral fits per 0.125 s show the pattern of temperature and
radius changes associated with the flux rising to the Eddington limit for
hydrogen poor material and then a decay at approximately constant
apparent radius after the
photosphere
has presumably returned to the surface of the neutron star. Taking the
flux at the start of this decay to give the observed Eddington limit at
the neutron star radius, the flux and color temperature run during the
decay can be fit to the relation calculated by Ebisuzaki and Nakamura
\cite{Ebisuzaki88} for the value of the effective temperature at the
Eddington limit. This temperature constrains the mass and radius:

\begin{equation}
kT_E = [(c/\sigma \kappa _o)(GM/R^2)]^{1/4} (1-2GM/Rc^2)^{3/8}
\end{equation}
Figure 5 shows the results.

\begin{figure}[htbp]
\parbox{75mm}{\epsfig{angle=0.,file=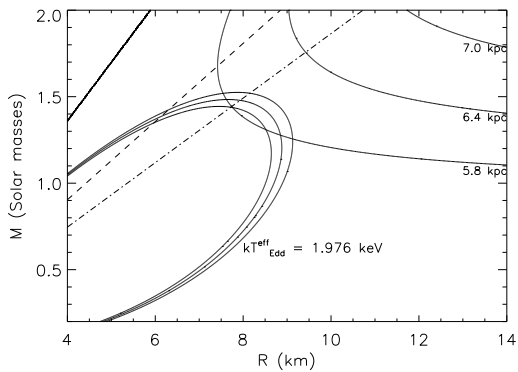,height=54mm}}
\caption{Mass versus radius constraints derived from a radius expansion burst
from 4U 1820-30. The set of three curves rising from the lower left corner
are derived by fitting for the Eddington effective temperature using the 
observed blackbody parameters and the theoretical pure helium atmosphere 
calculations of Ebisuzaki and Nakamura (1988).The dot-dash line from the
compactness limit constrains the values to its right.}
\label{fig:toosmall}
\end{figure}

\section{CONCLUSIONS}

Many, but not all bursts from some sources show possible coherent
strong oscillations. 
A correlary  is that bursts that are observed without oscillations do
not imply  that oscillations will
not be seen on other occasions. The period range of $3-1.7$ ms 
is about right for these sources eventually
becoming some of the recycled millisecond radio pulsars. The behavior at
onset of the burst suggests probably a single expanding hot spot. In all
cases so far analyzed, the burst tail frequency is no more than a couple
of Hz different from the rise frequency, so that the geometry of the
emission region must remain essentially the same. Spectral analysis
indicates the parameter varying is the average temperature in view. 
Coherence during the burst and the rationale for the small frequency changes
needs further study to answer outstanding questions.
The possibility of using burst spectra and the oscillations to constrain
the mass, radius, composition, and distance of the neutron stars holds great
promise.

\end{document}